\documentclass[runningheads]{llncs}
\usepackage[T1]{fontenc}
\usepackage{graphicx}
\usepackage{amsmath}
\usepackage{amstext}
\usepackage{amsfonts}
\usepackage[usestackEOL]{stackengine}
\usepackage{booktabs,multirow,makecell}

\def\mr[#1]#2#3{\multirowcell{#2}[#1]{#3}}
\begin{document}
\title{GACE: Learning Graph-Based Cross-Page Ads Embedding For Click-Through Rate Prediction}
\titlerunning{GACE: Learning Graph-Based Cross-Page Ads Embedding For CTR}
%
\author{Haowen Wang\inst{1} \and
Yuliang Du\inst{1} \and
Congyun Jin\inst{1} \and
Yujiao Li\inst{1} \and
Yingbo Wang\inst{1} \and
Tao Sun\inst{1} \and
Piqi Qin\inst{1} \and
Cong Fan\inst{1}}
\authorrunning{H. Author et al.}
%
\institute{
AntGroup\\
\email{\{wanghaowen.whw, duyuliang.dyl, jincongyun.jcy, 	liyujiao.lyj, wangyingbo.wyb, suntao.sun, piqi.qpq, fancong.fan\}@antgroup.com}}

\maketitle              
\begin{abstract}
Predicting click-through rate (CTR) is the core task of many ads online recommendation systems, which helps improve user experience and increase platform revenue. In this type of recommendation system, we often encounter two main problems: the joint usage of multi-page historical advertising data and the cold start of new ads. In this paper, we proposed GACE, a graph-based cross-page ads embedding generation method. It can warm up and generate the representation embedding of cold-start and existing ads across various pages. Specifically, we carefully build linkages and a weighted undirected graph model considering semantic and page-type attributes to guide the direction of feature fusion and generation. We designed a variational auto-encoding task as pre-training module and generated embedding representations for new and old ads based on this task. The results evaluated in the public dataset AliEC from RecBole and the real-world industry dataset from Alipay show that our GACE method is significantly superior to the SOTA method. In the online A/B test, the click-through rate on three real-world pages from Alipay has increased by 3.6\%, 2.13\%, and 3.02\%, respectively. Especially in the cold-start task, the CTR increased by 9.96\%, 7.51\%, and 8.97\%, respectively.

\end{abstract}
\section{Introduction}
With the increase of APP pages and advertising frequency, there are some challenges in improving the efficiency of essential advertising CTR tasks of e-commerce platform\cite{lipmaa2000ctr,chen2016deep} applications: For e-commercial platforms such as Taobao, Alipay, etc., we have recently seen an increase in recommendation channels for APPs on different pages. This means that the recommendation behavior and user interaction history information on multiple pages can be considered jointly. The recommendation performance on multiple pages can be improved by using the interaction information of multiple pages, including improving cold start advertising distribution efficiency.

Over the years, deep-learning-based models\cite{cheng2016wide,wang2017deep,guo2017deepfm,bert4rec_ref} have been proven to improve the efficiency of ads distribution due to their powerful ability of feature intersection and fusion. 
Despite the remarkable success of these models themselves, their performance still depends primarily on the input of embedded vectors in practical applications. A high-quality ad embedding vector has been proven to improve the accuracy of CTR's prediction effectively\cite{okura2017embedding,ouyang2019deep,zhou2018deep}.

In this paper, We believe differences between pages should be described and considered in a cross-page universal recommendation system. This concept should be applied to the item embedding generation process, which existing item embedding generation methods have yet to be paid attention to. We propose a graph-based cross-page ad embedding learning framework (GACE) from different perspectives, which is an improved variational graph auto-encoder\cite{kipf2016variational}. One ad's features are mainly composed of Semantic Knowledge (advertising text), Page Knowledge, and User Interaction Knowledge. We first designed a weighted undirected graph network\cite{pettie2005shortest} that obtains links based on the semantic similarity of advertising texts and the similarity of page representations. Based on the graph attention network \cite{velivckovic2017graph}, we precisely designed an auto-encoding task\cite{khawar2020learning} of an undirected weighted graph as the pre-training module, in which the variational graph attention encoder can adaptively extract information. Through such pre-training tasks, old and new ads can obtain embeddings considering cross-page neighbor information. The main contributions of this work can be summarized as follows:

\begin{itemize}
\item We build linkages based on the text information of the advertising content and page features to generate a weighted undirected graph to guide the process of ads information transfer.
\item We designed the pre-training task based on the improved 
variational graph auto-encoder for the weighted undirected graph of the ad, which can generate the embedding considering the neighbor information for either the new or old ad. 
\item We have conducted extensive experiments on public large-scale real-world datasets AliEC and offline experimental datasets from Alipay and evaluated through online A/B testing of the Alipay platform, which proves that GACE has achieved significant CTR improvement on each page, especially in the advertising cold start scenario.
\end{itemize}

\section{RELATED WORKS}
Our work is committed to improving the click-through rate of the recommendation system by generating ad item embedding under the consideration of cross-page information. It mainly involves several research fields: General CTR prediction of recommendation systems and learning of advertisement embedding.

\subsection{General CTR recommendation systems}
The CTR prediction task in online advertising recommendation systems is to predict the click probability of a given user for a given advertisement, which plays an increasingly important role in various personalized online advertising recommendation services. In recent years, deep learning methods based on deep neural networks can interact and represent features in complex ways, and a series of deep neural network models have been proposed. The Wide\&Deep\cite{cheng2016wide} model attempts to capture high-level feature interactions through a simple multilayer perceptron (MLP)\cite{2013Multilayer} network. DeepFM\cite{guo2017deepfm} and DCN \cite{wang2017deep} are proposed to handle complex feature interactions based on product operations. AutoInt\cite{2019AutoInt} uses the Multi-head Self-Attention mechanism to produce high-level composite features. In addition, in the scenario with sequence characteristics, a series of neural networks based on sequence characteristics are proposed: Deep Interest Network (DIN)\cite{zhou2018deep}, Behavior Sequence Transformer (BST)\cite{2019Behavior}, Deep Session Interest Network (DSIN)\cite{2019Deep} and Bert4Rec\cite{bert4rec_ref} etc.

However, the performance of these models depends mainly on the quality of the input item and user embedding and cannot effectively solve the cold start problem of new advertisements. They often fail to perform satisfactorily in the sparse page scenario or when distributing ads not in the training set. 

\subsection{Item Embedding Generation}
For item embedding, the first concept is to establish a one-hot vector for item id representation. The embedding learning methods based on collaborative filtering and matrix decomposition were subsequently proposed, but these methods have weak generalization ability and poor performance for items with sparse historical behavior. Researchers try to extend the concept of word embedding in the field of natural language process (NLP)\cite{2020Fundamentals} to the recommendation field, and propose Item2Vec\cite{2016Item2Vec} based on the concept of Word2Vec \cite{CHURCH2017Word2Vec}. With the development of graph neural networks, embedding learning methods based on graph networks have also been proposed, such as DeepWalk\cite{2014DeepWalk}, GraphSage\cite{2017Inductive}, etc. However, new advertising embedding without historical interaction cannot be learned via these methods. In particular, for the embedded learning of cold-start new items, our standard solution is to put forward a series of neighbor embedding methods based on manifold learning. Some researchers have put forward some generative methods from the perspective of meta-learning, such as the MetaEmbedding model\cite{2020Domain}, autodis\cite{2020AutoDis}, etc. However, these methods only play a role in learning the embedded representation of new advertisements, and it is difficult to improve the recommendation efficiency of existing old items.

\section{PRELIMINARIES}
\begin{figure}
\includegraphics[width=\textwidth]{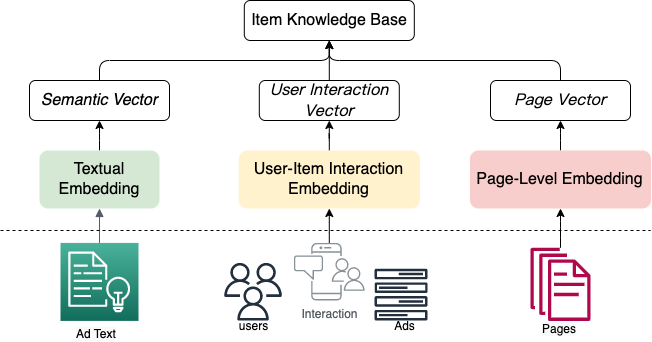}
\caption{Item Knowledge Base: semantic knowledge, user interaction knowledge and page knowledge} \label{fig1}
\end{figure}

The information stored in the item knowledge base mainly includes three parts as shown in Fig~\ref{fig1}: semantic knowledge, user interaction knowledge, and page knowledge. 

\subsection{Item Knowledge Base}
\subsubsection{Semantic Knowledge} Semantic knowledge is the content information of ads. Advertising recommendations usually include advertising text and corresponding attributes (such as font size, style, color, language, etc.)
\subsubsection{User Interaction Knowledge} User interaction knowledge is a vital part of the dynamic attributes of advertising. Its indicators (UV, PV, UVCTR, PVCTR) can reflect the transformation effect of advertising history and user satisfaction.

\textbf{UV: } unique visitors, the number of distinct individuals visiting an ad within the specified period (usually one day).

\textbf{PV: } page views, the number of times a specific ad is accessed in a specific period (usually one day).

\textbf{UVCTR: } click-through rate in UV, the percentage of unique people who see your ads and click it. The formula of UVCTR is clicks of unique visitors divided by impressions of unique visitors.

\textbf{PVCTR: } click-through rate in PV, the percentage of page views who see your ad and click it. The formula of PVCTR is clicks of ads divided by impressions of ads views.

\subsubsection{Page Knowledge} Page knowledge includes the identification of the page where this ad is distributed. We count the number of ads distributed on specific page channels and aggregate the average value of user interaction knowledge (UV, PV, UVCTR, PVCTR) of historical ads distributed on a specific page channel to serve as the embedded representation of the page.

\subsection{Problem Formulation}
The task of an online advertising system is to establish a prediction model to estimate the click probability of a specific user for a specific ad. Each instance includes multiple-field information: user information ('User ID', 'City', 'Age', etc.), item knowledge base, combined with the label from historical user interaction feedback. 

\section{METHODOLOGY}
\subsection{Overview}
We proposed a graph-based cross-page ads embedding learning method (GACE) by fully exploiting useful information in the item knowledge base. Our model contains two steps: graph creation and pre-training embedding learning. In the graph creation step, three entities' embedding vectors of each item are extracted or encoded from the item knowledge base, representing latent knowledge in each domain. Then we established a weighted undirected graph structure based on the semantic knowledge and page knowledge and set the splicing of entities embeddings in the item knowledge base as the initial embedding for the graph node.
In the pre-training embedding learning step, we proposed a pre-training task based on an improved graph auto-encoder for the weighted undirected graph established in the first step to learn the potential embedding representation of each advertising node in the graph. Accordingly, we obtain the optimization parameters of the ad embedding encoder and the potential vector representation of ad items. The graph creation and pre-training embedding learning steps will be further discussed in sections 4.2 and 4.3.

\subsection{Graph Creation}
\begin{figure}
\centering
\includegraphics[width=\textwidth/2]{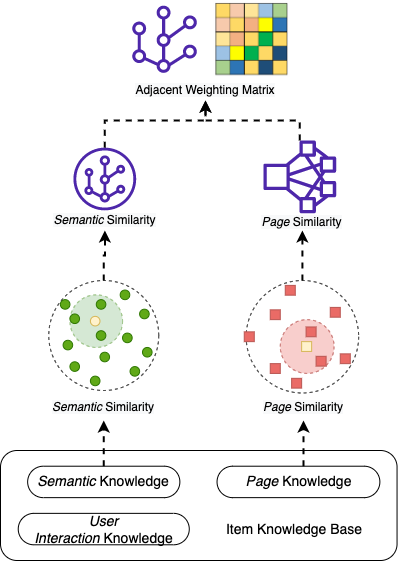}
\caption{The graph creation part for GACE using item knowledge base} \label{fig2}
\end{figure}

Typically, for a graph $\mathcal{G=(V,E)}$, where $\mathcal{V}=\big\{v_1,\ldots,v_n\big\}$ is a set of ad nodes and $\mathcal{E}$ is a set of edges with weightings, we combined and set the splicing of three entities' embeddings in the item knowledge base as the ad node's initial vector $v$. We proposed an adjacent weighting matrix $\mathbf{A} \in \mathbb{R}^{n \times n}$, where $\mathbf{A}_{i,j} \geq 0$ to represent the graph structure. If  $\mathbf{A}_{i,j}>0$, then there is an edge between ads item $i$ and ads item $j$, where $\mathbf{A}_{i,j}$ represents the edge weighting. If $\mathbf{A}_{i,j}=0$, then there is no connection between ads item $i$ and ads item $j$. The Adjacent weighting matrix is calculated based on semantic knowledge and page knowledge. Since the semantic knowledge of advertising is mainly composed of sentences, here we use the output of the sentence transformer\cite{devika2021deep} as the semantic knowledge vectors $\mathbf{s} \in \mathbb{R}^{k}$. As mentioned in section 2.1.3, we count the total number of ads placed on specific page channels and aggregate the average value of user interaction knowledge of historical ads placed on a specific page channel as Page knowledge vectors $\mathbf{s} \in \mathbb{R}^{d}$. The graph creation process is shown in Fig~\ref{fig2}.

We use dot product as the semantic knowledge similarity $\alpha$ and page similarity $\beta$, and calculate the advertising weight matrix $\mathbf{A}$ based on $\alpha$ and $\beta$ as follows:
\begin{equation}
   \alpha_{i,j} = s_i\cdot s_j
\end{equation}
\begin{equation}
    \beta_{i,j} = p_i\cdot p_j
\end{equation}
\begin{equation}
    ReLU(x) = max(0,x)
\end{equation}
\begin{equation}
    \ \mathbf{A}_{i,j} = ReLU(\alpha_{i,j})\cdot ReLU(\beta_{i,j})
\end{equation}

where $\alpha_(i,j)$ refers semantic knowledge similarity between ads item $i$ and ads item $j$, $s_i$ and $s_j$ are semantic knowledge vectors for ads item $i$ and ads item $j$. Similarly, $\beta_(i,j)$ is page similarity between ads item $i$ and ads item $j$, $p_i$ and $p_j$ are Page knowledge vectors for ads item $i$ and ads item $j$. $ReLU$ is a non-linear activation function.

\subsection{Pre-training Embedding Learning}
\begin{figure}
\includegraphics[width=\textwidth]{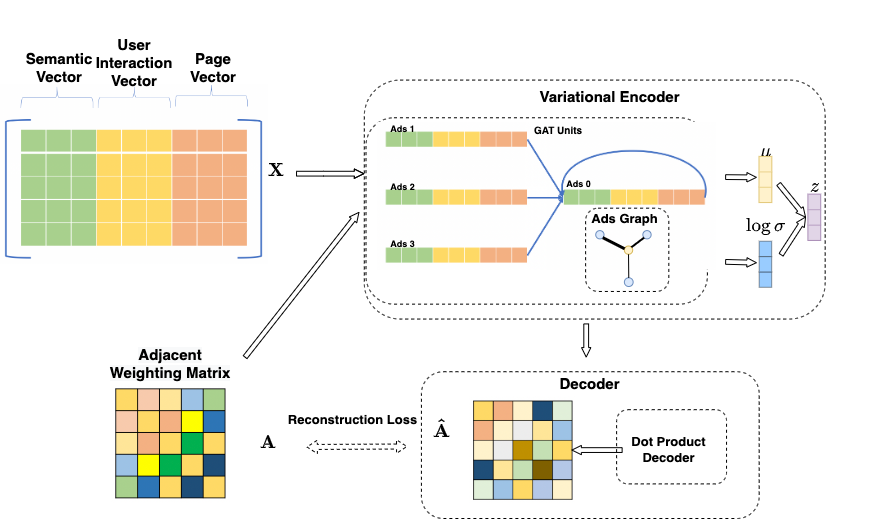}
\caption{The workflow for self-encode task based on variational graph auto-encoder.} \label{fig3}
\end{figure}

In this section, to integrate the knowledge of different entities of ads nodes and their neighbors, we designed a variational graph auto-encoder for an elaborately designed self-encode task as a pre-training module to recover the undirected weighting graph structure, as shown in Fig~\ref{fig3}. After the pre-training task, the encoder can generate embedding representations $Z$ for new and old ads.

\textbf{Encoder}: We designed an encoder based on variational encoder and graph attention neural network, which can adaptively adjust the weighting between ads nodes. Latent nodes vector $z_i$ for ads item $i$ are further introduced as:
\begin{equation}
   q(z_i\mid X, A) = \mathcal{N}(z_i \mid \mu_i, diag(\sigma_i^2)
\end{equation}
\begin{equation}
   GAT(A, X) = \gamma_{i,i} \mathbf{W}x_i + \textstyle \sum_{j\in\mathcal{N}(i)}\gamma_{i,j} \mathbf{W}x_i
\end{equation}
\begin{equation}
    \mathbf{\mu} =GAT_\mu(A,X)
\end{equation}
\begin{equation}
    \text{log} \mathbf{\sigma} =GAT_\sigma(A,X)
\end{equation}

where $x_i \in \mathbb{R}^{F}$. is the node vector of ads item $i$, and $F$ is the dimension of each node feature. $\mathbf{W} \in \mathbb{R}^{F'\times F}$ is the trainable weighting matrix to distill useful information, and $F'$ is the transformation size. $\mathcal{N}(i)$ is the set of neighbors of node $i$, and the cross-page attention weighting $\gamma_{i,j}$ is calculated as: 

\begin{equation}
   \gamma_{i,j} = \frac{exp(LeakyReLU(\alpha^\top[\mathbf{W}x_i||\mathbf{W}x_j])+\mathbf{\Theta}_{i,j})}{\sum_{k\in\mathcal{N}(i)}(exp(LeakyReLU(\alpha^\top[\mathbf{W}x_i||\mathbf{W}x_k])+\mathbf{\Theta}_{i,k})}
\end{equation}
\begin{equation}
   \mathbf{\Theta}_{i,j} =softmax_j(\mathbf{A}_{i,j})=\frac{exp(\mathbf{A}_{i,j})}{\sum_{k\in\mathcal{N}(i)}exp(\mathbf{A}_{i,k})}
\end{equation}
\begin{equation}
    LeakyReLU(x) =max(0,x)+negative\_slope\cdot min(0,x)
\end{equation}

where $LeakyReLU$ and $softmax$ are non-linear activation functions. A feed-forward neural network is proposed here and is parameterized by a weight vector $a \in \mathbb{R}^{2F'}$. $\mathbf{\Theta}_{i,j}$ is the enhanced and normalized edge weighting between ads item $i$ and ads item $j$.

\textbf{Decoder}: For non-probabilistic variant of the GAE model, we reconstruct the adjacent weighting matrix $\mathbf{\hat{A}}$ as:
\begin{equation}
    \mathbf{\hat{A}} = \sigma(ZZ^\top)
\end{equation}

where $\sigma(\cdot)$ is the ReLU non-linear activation function.

\textbf{Learning}: For GACE pre-training, we designed $\mathcal{L}_r$ to optimize the reconstruction of the adjacent weighting structure, and $\mathcal{L}_n$ ensures that the generated vector follows the normal distribution.
\begin{equation}
\mathcal{L} =\mathcal{L}_r+\mathcal{L}_n = \sum_{i=0}^nKL[q(\hat{\mathbf{A}_i})||p(\mathbf{A}_i)] + (-KL[q(z|X,\mathbf{A})||p(Z)])
\end{equation}

where $KL\lbrack q(\cdot)||p(\cdot)\rbrack$ is the Kullback-Leibler divergence\cite{joyce2011kullback} between distribution $q(\cdot)$ and $p(\cdot)$. We take \begin{math}p(Z)=\prod\nolimits_ip(z_i) \end{math} as a Gaussian prior, and take Kullback-Leibler divergence for $\mathcal{L}_r$ as the reconstruction loss to retain the weighting distribution information. We perform full-batch gradient descent\cite{soodabeh2020learning} and use the reparameterization trick\cite{wilson2017reparameterization} for training.
\section{Experiment}
\subsection{Dataset}
Considering cross-page information is significant, we elaborately selected the public datasets that have page-level information. Experiments are conducted on the public dataset AliEC\cite{dataset_aliec} 
 from RecBole\cite{recbole[2.0]} which has two pages (regarded as different resources PageID (PID) location) and a real-world CTR prediction dataset collected from three pages of Alipay.
\begin{table}[h]
\scriptsize
\centering
\caption{Statistics of RecBole Public Dataset AliEC and Alipay Real-World Offline Experimental Dataset}
\label{tab:data_description}
\tabcolsep=0.11cm
\centering
\begin{tabular}{cccccc}
\toprule
\textbf{DataSet} & \multicolumn{2}{c}{\textbf{AliEC}} & \multicolumn{3}{c}{\begin{tabular}[c]{@{}c@{}}\textbf{Alipay Real-World} \\ \textbf{Offline Experimental Dataset}\end{tabular}} \\  \cmidrule(lr){1-1}  \cmidrule(lr){2-3} \cmidrule(lr){4-6}
Page & Page1 & Page2 & Page1 & Page2 & Page3 \\
\midrule
Old Ads Num & 579323 & 752454 & 6238 & 4575 & 4306 \\
User Num & 403042 & 726392 & 1681441 & 2179180 & 5971524 \\
Samples for Warming Embedding and Training & 8760940 & 14235512 & 16082049 & 22217967 & 7776079 \\
Sample for Old Ads Testing & 1324123 & 2237386 & 4018443 & 5554002 & 1943008 \\
\midrule
New Ads Num & - & - & 610 & 415 & 407 \\
Samples for New Ads Testing & - & - & 6885571 & 4316894 & 7835890\\
 \bottomrule
\end{tabular}
\end{table}
 We designed offline comparison experiments, evaluated performance on different pages respectively, and then conducted an online A/B test. The statistics of experimental datasets are shown in Table~\ref{tab:data_description}

\subsection{Experimental Settings}
\subsubsection{Main CTR Prediction Models}Because GACE is the pre-training model for item ID embedding generation. They can be applied to various CTR prediction models which require item embeddings. We conduct experiments on the real-world datasets on the following base CTR prediction models: 

\textbf{Wide\&Deep: } Wide\&Deep Model has been widely accepted in industrial applications. It combined joint training with a wide DNN (for memorization) and a deep DNN (for generalization). We follow the practice in \cite{guo2017deepfm} to take the cross-product of user behaviors and candidates as wide inputs.

\textbf{DCN: } Deep\&Cross Model. It combined cross layers with DNN in Wide\&Deep.

\textbf{DeepFM: } DeepFM Model. It applies factorization machines as "wide" DNN module in Wide\&Deep.

\textbf{Bert4Rec: } Bert4Rec Model. It employs deep bidirectional self-attention to model user behavior sequences and is combined with DNN to process user and item candidates' features.

\subsubsection{Item ID Embedding Models} For each CTR prediction model, we evaluate the following embedding models, which generate initial embeddings for new and old ads.

\textbf{RndEmb: } It uses randomly generated embedding for new ads and looks up item knowledge base for old ads representation.

\textbf{NgbEmb: } It aggregates the initial embedding of neighbor items and generates the ads embedding $z\_ngb$ as follows: 
\begin{equation}
    z\_ngb_i=x_i+\sum\nolimits_{j\in\mathcal{N}(i)}x_j
\end{equation}

NgbEmb serves as a baseline that only considers neighbor information.

\textbf{N2VEmb: } Node2vec is a graph embedding method that comprehensively considers DFS and BFS neighborhoods. It can be regarded as an extension of deepwalk, which combines DFS and BFS random walk.

\textbf{GACEEmb: } Graph-based cross-page ad embedding. It generates ad item embedding for both new and old ads considering the cross-page and semantic knowledge.

\subsubsection{Parameter Setting} We set the dimension of the ads embedding as 15. The MLP part of Wide\&Deep, DeepFM, DCN and Bert4Rec is implemented using the same architecture: two-layer fully connected layers with [256, 128, 64] hidden units. For all attention layers in the above models, the number of hidden units is set to 128. All models are implemented in TensorFlow and optimized using AdamW optimizer. The grid-search strategy is applied to find the optimal hyper-parameters such as batch size (among 256, 512, 1024) and learning rate (1e-2, 1e-3, 5e-4, 1e-4). The experiment of each model with the optimal hyper-parameters is conducted for three times and the average result is reported. 

\subsubsection{Evaluation Scheme} Model performance is evaluated on different pages and model performance between new and old ads is also compared. We used the following evaluation matrix in our recommendation tasks. AUC and Loss are adopted for offline comparison experiments to evaluate models' performance.

\textbf{AUC: } AUC measures the entire two-dimensional area underneath the ROC curve. It is widely used for evaluating the classification model. It reflects the probability that the model will rank a randomly chosen positive example more highly than a randomly chosen negative example. Higher AUC indicates better model performance.

\textbf{Loss: } We use the cross-entropy loss for the click-through rate task on the test set to evaluate the learning process. 

Smaller loss indicates better model performance.
\begin{equation}
    loss=\frac{1}{Y}\sum\nolimits_{\mathcal{Y}\in Y} \lbrack-\mathcal{Y} \text{log} \hat{\mathcal{Y}} - (1-\mathcal{Y}) \text{log} (1-\hat{\mathcal{Y}})\rbrack
\end{equation}
\subsection{Result from model comparison on public AliEC Dataset}
In this section, we conduct model comparison experiments on the public AliEC Dataset and evaluate model performance on two different page scenarios.

Table~\ref{tab:results_for_public_private} shows the results from the public AliEC Dataset. Models with deep feature interactions perform better than the original wide\&deep models. GACEEmb stands out significantly among all the other item-embedding competitors, which shows in the two pages evaluated results respectively. We owe this to the neighbor information aggregation generation mechanism considering the similar content and similar distribution pages. Through improved variational graph auto-encoding, GACE obtained the enhanced representation of ad items. 

\begin{table}[h]
\footnotesize
\centering
\caption{Test AUC and loss for Public Dataset Aliec on \textbf{Existing old ads}. Pred Model: CTR Prediction Model. Emb Model: Embedding Generation Model.}
\label{tab:results_for_public_private}
\tabcolsep=0.11cm
\centering
\begin{tabular}{cccccc}
\toprule
\multirow{2}{*}{\begin{tabular}[c]{@{}c@{}}\bfseries Pred\\ \bfseries Model\end{tabular}} & \multirow{2}{*}{\begin{tabular}[c]{@{}c@{}}\bfseries Embed\\ \bfseries Model\end{tabular}} 
 & \multicolumn{2}{c}{Page1} & \multicolumn{2}{c}{Page2}  \\ \cmidrule(lr){3-4}  \cmidrule(lr){5-6}  
 &  & AUC & Loss & AUC & Loss \\
 \midrule
     \mr[-0.5ex]{4}{Wide \& Deep} 
    & RndEmb & 0.520 &	0.511 &	0.505 &	0.640 \\
    & NgbEmb & 0.537 &	0.489 &	0.537 &	0.562 \\
    & N2VEmb & 0.578 &	0.453 &	0.590 &	0.428 \\
    & GACEEmb & \textbf{0.588} & \textbf{0.346} & \textbf{0.595} & \textbf{0.337} \\
   \midrule
    \mr[-0.5ex]{4}{DCN} 
    & RndEmb & 0.573 &	0.352 &	0.511 &	0.290  \\ 
    & NgbEmb & 0.578 & 0.339 & 0.570 & 0.272 \\
    & N2VEmb & 0.586 & 0.328 &	0.597 &	0.255 \\
    & GACEEmb & \textbf{0.595} & \textbf{0.315} & \textbf{0.602} & \textbf{0.252} \\
   \midrule
    \mr[-0.5ex]{4}{DeepFM} 
    & RndEmb & 0.557 &	0.345 &	0.572 &	0.295 \\ 
    & NgbEmb & 0.582 & 0.314 & 0.600 & 0.260 \\
    & N2VEmb & 0.593 & 0.292 &	0.605 &	0.257  \\
    & GACEEmb & \textbf{0.599} & \textbf{0.278} &	\textbf{0.616} &	\textbf{0.252} \\
    \midrule
    \mr[-0.5ex]{4}{Bert4Rec} 
    & RndEmb & 0.579 &	0.318 &	0.559 &	0.282 \\ 
    & NgbEmb & 0.591 &	0.309 &	0.590 &	0.261 \\
    & N2VEmb & 0.602 & 0.299 & 0.608 & 0.254 \\
    & GACEEmb & \textbf{0.610} & \textbf{0.258} & \textbf{0.627} &	\textbf{0.245} \\
 \bottomrule
\end{tabular}
\end{table}

\subsection{Result from model comparison on Alipay offline experimental Dataset}
In this section, we conducted offline comparison experiments and evaluate model performance on different Alipay pages respectively. Typically, we evaluated the impact of the GACE embedding generation framework on the recommendation effect of new and old ads in different CTR prediction models.  

\subsubsection{Effectiveness of cold-start ads} 
Table~\ref{tab:results_for_private_cold} exhibits the performance of various embedding models based on different CTR prediction models for cold-start ads. It can be observed that NgbEmb performs better than RndEmb, which means that neighbors’ related attributes can contribute useful information and alleviate the cold-start problem. N2VEmb improves better than NgbEmb, indicating that just considering the simple average pre-training neighbor ID embedding is insufficient, and considering more neighbors' information is proved effective. GACE leads to a more stable and effective performance improvement than NgbEmb and N2VEmb. It shows that the marginal effect of aggregating neighbor information can be effectively improved by considering the content and page information.

\begin{table}[h]
\footnotesize
\centering
\caption{Test AUC and loss for Alipay Real-World Dataset on \textbf{Cold-start ads}. Pred Model: CTR Prediction Model. Emb Model: Embedding Generation Model.}
\label{tab:results_for_private_cold}
\tabcolsep=0.11cm
\centering
\begin{tabular}{cccccccc}
 \toprule
\multirow{2}{*}{\begin{tabular}[c]{@{}c@{}}\bfseries Pred\\ \bfseries Model\end{tabular}} & \multirow{2}{*}{\begin{tabular}[c]{@{}c@{}}\bfseries Embed\\ \bfseries Model\end{tabular}} 
 & \multicolumn{2}{c}{Page1} & \multicolumn{2}{c}{Page2}  & \multicolumn{2}{c}{Page3} \\ \cmidrule(lr){3-4}  \cmidrule(lr){5-6}  \cmidrule(lr){7-8} 
 &  & AUC & Loss & AUC & Loss & AUC & Loss\\
 \midrule
     \mr[-0.5ex]{4}{Wide \& Deep} 
    & RndEmb &  0.700 & 1.074 & 0.531 & 0.339 & 0.344 & 0.997 \\
    & NgbEmb &  0.733 & 0.817 & 0.551 & 0.331 & 0.463 & 0.449 \\
    & N2VEmb &  0.741 & 0.759 & 0.566 & 0.326 & 0.541 & 0.386 \\
    & GACEEmb & \textbf{0.790} & \textbf{0.645} & \textbf{0.608} & \textbf{0.264} & \textbf{0.892} & \textbf{0.247} \\
   \midrule
    \mr[-0.5ex]{4}{DCN} 
    & RndEmb & 0.610 & 1.185 & 0.569 & 0.343 & 0.345 & 0.873  \\ 
    & NgbEmb & 0.641 & 1.133 & 0.597 & 0.314 & 0.446 & 0.467  \\
    & N2VEmb & 0.702 & 0.868 & 0.609 & 0.295 & 0.511 & 0.429 \\
    & GACEEmb & \textbf{0.789} & \textbf{0.594} & \textbf{0.635} & \textbf{0.279} & \textbf{0.706} & \textbf{0.380} \\
   \midrule
    \mr[-0.5ex]{4}{DeepFM} 
    & RndEmb & 0.557 &	0.727 & 0.764 & 0.564 & 0.416 & 0.331 \\ 
    & NgbEmb & 0.582 & 0.759 & 0.854 & 0.611 & 0.351 & 0.445 \\
    & N2VEmb & 0.593 & 0.772 & 0.691 & 0.649 & 0.294 & 0.513 \\
    & GACEEmb & \textbf{0.796} & \textbf{0.536} & \textbf{0.663} & \textbf{0.273} & \textbf{0.874} & \textbf{0.251}  \\
    \midrule
    \mr[-0.5ex]{4}{Bert4Rec} 
    & RndEmb & 0.744 & 0.791 & 0.565 & 0.417 & 0.380 & 1.125  \\ 
    & NgbEmb & 0.758 & 0.787 & 0.593 & 0.335 & 0.428 & 0.397  \\
    & N2VEmb & 0.764 & 0.727 & 0.613 & 0.292 & 0.557 & 0.371 \\
    & GACEEmb & \textbf{0.783} & \textbf{0.548} & \textbf{0.676} & \textbf{0.244} & \textbf{0.911} & \textbf{0.225} \\
 \bottomrule
\end{tabular}
\end{table}

\subsubsection{Effectiveness of old ads} Various embedding models' performances for old ads embedding warming up are compared in Table~\ref{tab:results_for_private_old}. The item warm-up graph contains both old and new ads so that we can provide the embeddings of old ads' from GACE learning. It is observed that GACE performs best in the cold-start phase and the warm-up phase, indicating that aggregating information considering the physical and semantic neighbors can effectively improve old ads' knowledge representation. The pre-training loss design with double KL divergence helps the embedded generation in the pre-training retain the information between ads to the maximum extent and map it to a higher space of information expression.

\begin{table}[h]
\footnotesize
\centering
\caption{Test AUC and loss for Alipay Real-World Dataset on \textbf{Existing old ads}. Pred Model: CTR Prediction Model. Emb Model: Embedding Generation Model.}
\label{tab:results_for_private_old}
\tabcolsep=0.11cm
\centering
\begin{tabular}{cccccccc}
 \toprule
\multirow{2}{*}{\begin{tabular}[c]{@{}c@{}}\bfseries Pred\\ \bfseries Model\end{tabular}} & \multirow{2}{*}{\begin{tabular}[c]{@{}c@{}}\bfseries Embed\\ \bfseries Model\end{tabular}} 
 & \multicolumn{2}{c}{Page1} & \multicolumn{2}{c}{Page2}  & \multicolumn{2}{c}{Page3} \\ \cmidrule(lr){3-4}  \cmidrule(lr){5-6}  \cmidrule(lr){7-8} 
 &  & AUC & Loss & AUC & Loss & AUC & Loss\\
 \midrule
     \mr[-0.5ex]{4}{Wide \& Deep} 
    & RndEmb &0.705 & 0.811 & 0.562 & 0.651 & 0.863 & 0.573\\
    & NgbEmb &0.727 & 0.776 & 0.598 & 0.572 & 0.881 & 0.508\\
    & N2VEmb &0.783 & 0.719 & 0.656 & 0.435 & 0.921 & 0.253\\
    & GACEEmb & \textbf{0.803} & \textbf{0.549} & \textbf{0.714} & \textbf{0.343} & \textbf{0.939} & \textbf{0.207} \\
   \midrule
    \mr[-0.5ex]{4}{DCN} 
    & RndEmb &0.796 & 0.558 & 0.621 & 0.295 & 0.903 & 0.215 \\ 
    & NgbEmb &0.802 & 0.537 & 0.693 & 0.277 & 0.912 & 0.156 \\
    & N2VEmb &  0.813 & 0.520 & 0.726 & 0.259 & 0.938 & 0.137 \\
    & GACEEmb &\textbf{0.826} & \textbf{0.499} & \textbf{0.731} & \textbf{0.256} & \textbf{0.952} & \textbf{0.112} \\
   \midrule
    \mr[-0.5ex]{4}{DeepFM} 
    & RndEmb &0.793 & 0.548 & 0.664 & 0.300 & 0.902 & 0.165 \\ 
    & NgbEmb &  0.829 & 0.498 & 0.697 & 0.265 & 0.911 & 0.156 \\
    & N2VEmb &0.844 & 0.463 & 0.702 & 0.261 & 0.945 & 0.133 \\
    & GACEEmb & \textbf{0.867} & \textbf{0.441} & \textbf{0.727} & \textbf{0.256} & \textbf{0.961} & \textbf{0.124} \\
    \midrule
    \mr[-0.5ex]{4}{Bert4Rec} 
    & RndEmb & 0.838 & 0.505 & 0.671 & 0.287 & 0.922 & 0.150 \\ 
    & NgbEmb & 0.856 & 0.490 & 0.708 & 0.266 & 0.939 & 0.144 \\
    & N2VEmb &0.872 & 0.475 & 0.730 & 0.258 & 0.946 & 0.125 \\
    & GACEEmb  &\textbf{0.883} & \textbf{0.409} & \textbf{0.753} & \textbf{0.249} & \textbf{0.963} & \textbf{0.112} \\
 \bottomrule
\end{tabular}
\end{table}

\subsection{Result from online A/B testing}
In order to address the limitations of offline evaluation and demonstrate the practical value of GACE, we further deployed the GACE model in an actual online environment, that owns tens of millions of users access it daily.

An online A/B test is designed to further evaluate the performance of GACE. We conducted a week-long rigorous A/B testing with three objectives: to cover more users, cover more cold start items, and collect more reliable results. Bert4Reg with N2VEmb is the main model already serving in the recommendation system. We allocated 10\% of Bert4Rec traffic with N2VEmb as the baseline and compared 10\% of Belt4Rec traffic with GACEEmb. We evaluated the new and old items that appeared this week separately. 

Table~\ref{tab:online} illustrates the cumulative relative improvement of the experimental model compared to the baseline model. The average CTR of using GACE has increased by 3.61\%, 2.13\%, and 3.02\% respectively. For cold-start ads distribution, the CTR has increased by 9.96\%, 7.51\%, and 8.97\% respectively. It is worth noting that both improvements have statistical significance (p-value less than 0.05). This practice-oriented experiment demonstrates the effectiveness of our model in real-world recommendation scenarios.

\begin{table}[h]
\footnotesize
\centering
\caption{Cumulative relative improvement of the experimental model compared to the baseline model for a week. Baseline: Bert4Rec + N2VEmb.}
\label{tab:online}
\tabcolsep=0.11cm
\centering
  \begin{tabular}{*{5}{l}}
   \toprule
   \textbf{Model} & \textbf{Eval scope} & \textbf{Page1} & \textbf{Page2} & \textbf{Page3}\\
   \midrule
   \mr[-0.5ex]{2}{Bert4Rec\\+GACEEmb} 
   & All ads &+3.61\% & +2.13\% & +3.02\% \\ 
   \cmidrule(lr){2-5}
   & Cold-start ads &+9.96\% & +7.51\% & +8.97\% \\
   \bottomrule
  \end{tabular}
\end{table}
\section{Conclusion}
This paper addresses the CTR prediction problem for cross-page ads whose ID embeddings still need well-learned. Graph-based cross-page ad embedding (GACE) can effectively learn how to generate desirable ad item embedding using cross-page data based on graph neural networks. It takes into account the page-level similarity relationship and semantic-based content relationship simultaneously, establishes a graph to connect all ads across pages, and adaptively extracts the information from cross-page adjacent ads. Because of the characteristics of the generative model, it can generate efficient embedding representations of new and old ads simultaneously. The experiment results show that GACE can effectively improve the CTR task performance for both new and old ads on four major deep-learning-based models. In the future, we will consider enhancing neighbor representation, try other methods to retrieve more information from neighbors and extend it to CVR and GMV Recall application scenarios.

%
%
%
\newpage
\bibliographystyle{splncs04}
\bibliography{sample-base}
%




\end{document}